\documentclass{article}

\usepackage{epsfig}
\usepackage{subfigure}
\usepackage{calc}
\usepackage{graphicx}

\usepackage{amsfonts}
\usepackage{amsmath}
\usepackage{amssymb}
\usepackage{latexsym}

\begin{document}

\title{An External Description for MIMO Systems Sampled
in an Aperiodic Way}
\date{}
\author{Amparo F\'{u}ster-Sabater\\
{\small Institute of Applied Physics, C.S.I.C.}\\
{\small Serrano 144, 28006 Madrid, Spain} \\
{\small amparo@iec.csic.es}}

\maketitle

\begin{abstract}

An external description for aperiodically sampled MIMO linear
systems has been developed. Emphasis is on the sampling period
sequence, included among the variables to be handled. The
computational procedure is simple and no use of polynomial matrix
theory is required. This input/output description is believed to
be a basic formulation for its later application to the problem of
optimal control and/or identification of linear dynamical systems.


\end{abstract}

\section{Introduction}
\footnotetext{Work supported by Ministerio de Educaci\'{o}n y
Ciencia (Spain).\\
First version published in IEEE Transactions on Automatic Control.
Volume 33, No. 4, pp. 381-384. April 1988.}

There are two different ways of describing dynamical systems:
\begin{description}
\item (i) by means of input/output relations;
\item (ii) by means of state variables. \end{description}

In the classical or frequency-domain approach, systems are
described by transfer functions which reflect just the external or
input/output properties of the system. However, this mode of
description entails some difficulties concerning stability and
realization \cite{Jury}, \cite{Ragazzini}.

The modern or time-domain approach turns around the axiomatic
concept of state. The method is exact in defining the notion of
dynamical systems and also describes all internal couplings among
the system variables \cite{Kalman}, \cite{Zadeh}. Nevertheless,
the procedure became somewhat disappointing due to the necessity
of finding state-variable models and to the implicit assumption
that all state-variables are accessible for direct measurement.
This assumption is justified in mechanical or electrical systems
but it is not generally satisfied for plants in chemical, gas,
paper, and other industries.

These considerations were responsible for the comeback of transfer
function methods \cite{Wolovich}-\cite{Kucera}.

On the other hand, the enormous increase in the use of digital
computers in process control has stimulated studies in the field
of discrete systems for both types of representation. See
\cite{Kailath}-\cite{Astrom} and also the above mentioned
references. All of them are concerned with constant sampling
period, which is convenient for the simplicity of implementation
and mathematical treatment. However, the general case of aperiodic
sampling is \textit{a priori} capable of more favorable solutions
to the problem of control and/or identification of dynamical
systems, and it is also feasible with modern time-sharing
equipment.

In this work, an input/output modeling technique for aperiodic
sampling linear systems has been developed. The external
description includes the sampling sequence among the variables to
be handled. The system is described by input/output data according
to the actual experimentation conditions. Although the
multivariable case is covered, the complexity of the polynomial
matrix theory is avoided.

The procedure is believed to be a basic formulation for its later
application to the synthesis of linear control systems sampled in
an aperiodic way, since most of these techniques for
nonperiodically sampled systems rely exclusively on the
state-space equations \cite{Nicoletti}-\cite{Troch}.

\section{Basic Assumptions}
Our discussion is restricted to the following:
\begin{description}
\item (\textit{a}) linear time-invariant multivariable
    dynamical systems of finite order;
\item (\textit{b}) systems whose transfer function is a $p
    \times m$ matrix (\textit{m}-inputs, \textit{p}-outputs),
    where the different entries are strictly proper rational
    functions.
\end{description}

We end this preliminary section with the following statement.

\textit{Statement:} Let $(G_l)$ be a
family of vector functions

\[G_l; |R^n \rightarrow  |R^n  \qquad G_l \in C^\infty(|R^n , |R^n )
\qquad (l=0,1, \ldots, n)
\]

$C^\infty(|R^n , |R^n )$ being the set of infinitely differentiable functions on $|R^n$. If the
following conditions are verified:
\begin{description}
\item (\textit{a}) there exists an integer $r \leq n$ such
    that the elements $(G_r(z))$ are linearly independent for
    all $z \in |R^n $

\item (\textit{b}) there exists an integer $k > r$ such that
    $(G_r(z))$ depends linearly on $(G_0(z),  \ldots$ $,
    G_r(z))$
\end{description}

\noindent then, there are functions $f_0, f_1, \ldots , f_n \in
C^\infty(|R^n , |R )$ such that the following expression holds:

\begin{equation}\label{eq:1}
\sum\limits_{l=0}^{n}f_{n-l}(z) \; G_l(z)=0  \qquad \forall z \in |R^n .
\end{equation}

The previous result is a direct consequence of the Cramer Rule; for more
details see \cite{Nomizu}.

\section{External Description for Nonperiodically Sampled Linear Systems}

\subsection{Input/Output Modeling Technique}
Let $H(s)$ be the matrix transfer function of a linear time-invariant
multivariable system.
\begin{equation}\label{eq:2}
H(s)=(H_{rq}(s)) \qquad (r=1, \ldots, p ), (q = 1, \ldots, m)
\end{equation}

\noindent let us rewrite $H(s)$ as
\begin{equation}\label{eq:3}
H(s)=\frac{N(s)} {d(s)}
\end{equation}
where

\begin{equation}\label{eq:4}
d(s)= s^n + d_1 s^{n-1}+ \ldots + d_n
\end{equation}

\noindent is the least common multiple of the denominators of the entries of $H(s)$.

In the time domain, the impulse response $h(t)$ can be written as

\begin{equation}\label{eq:5}
h(t)=(h_{rq}(t))=\left(\begin{array}{c}
                   h_1(t) \\
                   \vdots \\
                   h_p(t)
                 \end{array}\right)
\end{equation}

\noindent where the \textit{r}th row can also be written in matrix
form by means of the triad $(A, C, B_r)$. In fact,

\begin{equation}\label{eq:6}
h_r(t)=C\;exp(At)B_r \qquad (r=1, \ldots, p)
\end{equation}

with
$A$ = a bottom-companion matrix with last row

\begin{equation}\label{eq:7}
-[d_n, d_{n-1}, \ldots, d_1]
\end{equation}

\begin{equation}\label{eq:8}
C=(1,0, \ldots, 0)_{1 \times n}
\end{equation}

\begin{equation}\label{eq:9}
B_r= \left(
       \begin{array}{ccc}
         h_{r1}(0) & \ldots & h_{rm}(0) \\
         \dot{h}_{r1}(0)  & \ldots & \dot{h}_{rm}(0) \\
         \vdots & \ldots & \vdots \\
         h_{r1}^{(n-1)}(0) & \ldots & h_{rm}^{(n-1)}(0) \\
       \end{array}
     \right)_{n\times m} .
\end{equation}

Remark that the column vectors of $B_r$ correspond to the \textit{n}-first Markov
parameters of the scalar impulse responses $h_{rq}(t) \; (q = 1, \ldots, m)$.

It should be noticed that the triad $(A, C, B_r)$ leads us naturally to the
observability canonical realization from the vector impulse response
$h_r(t)$.

From this triad, we are going to define a family of vector functions
$G_l;$ $|R^n \rightarrow  |R^n \;\; (l = 0 , \ldots, n)$ given by

\begin{equation}\label{eq:10}
G_j(z)= C \; exp(A(z_1 + \ldots + z_j)) \qquad (j=1, \ldots, n )
\end{equation}
\begin{equation}\label{eq:11}
G_0(z)= C
\end{equation}

with

\begin{equation}\label{eq:12}
z=(z_1,  \ldots , z_n) \in |R^n .
\end{equation}

Thus, the vector impulse response $h_r(t)$ can be written in terms of these
functions as
\begin{equation}\label{eq:13}
h_r(z_1+\ldots+z_j)=G_j(z)B_r \qquad (r=1, \ldots, p).
\end{equation}

From an analytic viewpoint, the functions $G_l$ belong to $C^\infty(|R^n , |R^n )$
as composition of $C^\infty$ functions.

It has been proved \cite{Troch} that there is an open interval $I$ of $|R$ such that
the vectors $(G_0(z), G_1(z), \ldots, G_{n-1}(z))$ defined as before are linearly
independent for each $z \in I \times I \times \ldots \times I= I^n$.

In this case, it is easy to see that for the new domain $I^n$ the conditions a)
and b) in the previous statement hold. Hence, there will be functions $f_l(z)
\in C^\infty(I^n, |R) \; (l = 0, \ldots, n)$ such that

\begin{equation}\label{eq:14}
\sum\limits_{l=0}^{n}f_{n-l}(z) \; G_l(z)=0  \qquad \forall z \in I^n .
\end{equation}

Thus, the functions $(f_1(z), \ldots, f_{n}(z))$ can be obtained by solving a
compatible system of linear equations. (For simplicity $(f_0(z) = - 1)$.

In fact, the general form of these functions is

\begin{equation}\label{eq:15}
f_{n-l}(z)= \frac {Det(G_0^{'}(z), \ldots , G_{n}^{'}(z), \ldots, G_{n-1}^{'}(z))} {Det(G_0^{'}(z), \ldots , G_{n-1}^{'}(z))}  \qquad (l = 0 , \ldots, n-1)
\end{equation}

\noindent(' denotes the transpose) where the numerator is the determinant obtained
from the matrix $(G_0^{'}(z), \ldots , G_{n-1}^{'}(z))$ by replacing the $(l+1)$th column
by the column vector $G_n^{'}(z)$.

Now, we multiply both sides of (14) by

\begin{equation}\label{eq:16}
exp(Az^*) B_r \qquad (r=1, \ldots, p )
\end{equation}

with $z^*$ taking successively the values

\begin{equation}\label{eq:17}
z^*= -\sum\limits_{i=1}^{l} z_i \qquad (l=1, \ldots, n)
\end{equation}

and we get in each case

\begin{equation*}\label{eq:18a}
C \; exp(A(z_{l+1}+\ldots+z_n)) B_r
\end{equation*}
\begin{equation*}\label{eq:18b}
=\sum\limits_{i=1}^{n-l} f_i(z) C \;exp A(z_{l+1}+\ldots+z_{n-i}))B_r
\end{equation*}
\begin{equation}\label{eq:18c}
+ \sum\limits_{i=0}^{l-1} f_{n-i}(z) C \;exp( - A(z_{i+1}+\ldots+z_{l}))B_r
\end{equation}
\begin{equation*}
(l=1, \ldots, n), (r = l , \ldots, p).
\end{equation*}

Finally, we define

\begin{equation}\label{eq:19}
g_{n-l}^r(z)=\sum\limits_{i=0}^{l-1} f_{n-i}(z) C \;exp( - A(z_{i+1}+\ldots+z_{l})) B_r
\end{equation}
\begin{equation*}\label{eq:19}
= (g_{n-l}^{r1}(z), \ldots, g_{n-l}^{rm}(z)).
\end{equation*}
At this point, we can identify the components of $z$ with the elements of
the sampling period sequence.

In fact,

\begin{equation}\label{eq:20}
z_{n-l}=t_{k-l} - t_{k-l+1}=T_{k-l} \qquad (l=0, \ldots, n-1)
\end{equation} 

where $t_{k-l}$ are the sampling instants and $T_{k-l}$ the length of the sampling
intervals.

Thus, at an arbitrary sampling instant, say $t_k \;(k \geq n)$, we can condense
the preceding expressions into two sets of equations involving the
functions $f_i, g^{rq}$ and $h_{rq}$ as follows:

\begin{equation}\label{eq:21}
\sum\limits_{i=0}^{n} f_{i}h_{rq}(t_{k-i}-t_j) + g_{k-j}^{rq}=0 \qquad (j=k, \ldots, k-n+1)
\end{equation} 

\begin{equation}\label{eq:22}
\sum\limits_{i=0}^{n} f_{i}h_{rq}(t_{k-i}-t_j)=0 \qquad (j=k-n, \ldots, 0)
\end{equation}
\begin{equation*}
\qquad \qquad (r=1, \ldots, p), \,(q = l, \ldots, m).
\end{equation*} 

Note that, at time $t_k$, the functions $f_i, \,g^{rq}$ will depend on the sampling
interval lengths $(T_{k-n+1},\ldots,T_k)$ and so on.

Now, we multiply (21) and (22) by $(u_j^q) \; (q = 1, \ldots, m), (j = k,
k - 1, \ldots, 0)$, respectively, $(u_j^q$ being the \textit{q}th impulse input of the
system at the sampling instant $t_j$) and summing all these expressions, we
get

\begin{equation*}\label{eq:23a}
\sum\limits_{q=1}^{m} \big[ f_1\; \big(\sum\limits_{l=0}^{k-1} h_{rq}(t_{k-1}-t_l)u_l^q\big) + \ldots
\end{equation*}
\begin{equation}\label{eq:23b}
+ f_n\; \big(\sum\limits_{l=0}^{k-n} h_{rq}(t_{k-n}-t_l)u_l^q\big) + \sum\limits_{j=0}^{n-1}g_j^{rq}u_{k-j}^q\big]
\end{equation}
\begin{equation*}\label{eq:23c}
= \sum\limits_{q=1}^{m}  \sum\limits_{l=0}^{k}  h_{rq}(t_k-t_l)u_l^q.
\end{equation*}

Then, making use of the convolution expression

\begin{equation}\label{eq:24}
y_k^r= \sum\limits_{q=1}^{m}  \sum\limits_{l=0}^{k}  h_{rq}(t_k-t_l)u_l^q
\end{equation}

($y_k^r$ being the \textit{r}th output of the system at time $t_k$) the above expression
becomes
\begin{equation}\label{eq:25}
y_k^r=\sum\limits_{i=1}^{n}f_i y_{k-i}^r + \sum\limits_{q=1}^{m}  \sum\limits_{j=0}^{n-1}  g_{j}^{rq} u_{k-j}^q
\end{equation}

which is the input output description for linear time-invariant MIMO
systems sampled in an aperiodic way. Each system output at time $t_k$ can be
written as a linear combination of the same output and of the different
inputs at previous instants.
The expression (25) generalizes to the aperiodic case the well-known
input/output representation for linear systems sampled periodically.
The sequence of sampling intervals is implicit in the arguments of the
functions $f_i, \,g_j^{rq}$. Consequently, this freedom in the choice of the sampling
instants can be used in the solution of control problems, propagation of
measuring errors, parameter estimation, and related topics \cite{Nicoletti}-\cite{Fuster}.

It is convenient to note that the functions $f_i$ are the same for every
system output, while the functions $g_j^{rq}$ depend on the corresponding
impulse response $h_{rq}$.

\subsection{Simplified Computation of the Functions $f_i$, $g_j^{rq}$}
Companion matrices are an important example of nonderogatory
matrices, which have only one (normalized) eigenvector associated with
each distinct eigenvalue. This means that

\begin{description}
\item (i) the Jordan canonical form is clearly simplified
    (there is only one Jordan block for each distinct
    eigenvalue);
\item (ii) the similarity transformation of the given matrix
    to the Jordan form can be obtained in a standard way.
\end{description}

Thus, the computation of the Jordan canonical form for this kind of
matrix is quite easy.
Indeed,
\begin{equation}\label{eq:26}
A=TJT^{-1}
\end{equation}

where $J$ is the Jordan canonical form of the matrix $A$ and $T$ is an
invertible matrix of a well-known general form \cite{Kailath}.
In this way, (14) becomes
\begin{equation}\label{eq:27}
\sum\limits_{l=0}^{n}f_{n-l}(z)\, x_0 \, exp(J\alpha_l)=0
\end{equation}
with
\begin{equation}\label{eq:28}
\alpha_l= z_1+ \ldots + z_l \qquad (l=1, \ldots, n), (\alpha_0 = 0)
\end{equation}

\begin{equation}\label{eq:29}
x_0= CT.
\end{equation}

Let $(\varphi_l)\; (l = 1, \ldots, n)$ be the fundamental system of solutions of a \textit{n}th
order homogeneous linear differential equation whose characteristic
polynomial is $d(s)$.

In this case, factorizing $Det (x_0 \;exp(J\alpha_l))$ and cancelling common
factors in (15), the general form of the functions $f_i$ can be simplified to

\begin{equation}\label{eq:30}
f_i=\frac{\Delta_i} {\Delta} \qquad (i=1, \ldots, n)
\end{equation}
with

\begin{equation}\label{eq:31}
\Delta=Det(\varphi_{l}(\alpha_j)) \qquad (j=0, \ldots, n-1), (l = l , \ldots, n)
\end{equation}

and $\Delta_i$ is analogous but replacing the argument of the \textit{i}th column by $\alpha_n$.
The functions $g_j^{rq}$ can be easily computed from (21) for the new
simplified form of the functions $f_i$.

\subsection{Choice of the Sampling Sequence}
The procedure developed imposes nonrestrictive conditions on the
sampling sequence in order to guarantee the linear independence of the
vectors $(G_0(z), \ldots,$ $ G_{n-1}(z))$.

Strategies to determine the set $I^n \subset  |R^n$ whose elements $z$ verify the
above condition can be found in \cite{Troch} (in an analytic way) and in \cite{Fuster} (in a
geometric way).

\section{An Illustrative Example}
Let $H(s)$ be a two-input, two-output transfer function matrix
\begin{equation}\label{eq:32}
H(s)=\left(
  \begin{array}{cc}
   \frac{1} {s+1} & \frac{2} {s+1} \\
     \, &  \, \\
    \frac{1} {(s+1) (s+2)} & \frac{1} {s+2} \\
  \end{array}
\right)
\end{equation}

\begin{equation}\label{eq:33}
d(s)= s^2 + 3s + 2=(s+1) (s+2) \qquad (n=2)
\end{equation}

\begin{equation}\label{eq:34}
\varphi_1(t)=exp(-t)
\end{equation}

\begin{equation}\label{eq:35}
\varphi_2(t)=exp(-2t)
\end{equation}

The impulse response matrix will be

\begin{equation}\label{eq:36}
h(t)= \left(
        \begin{array}{cc}
          exp(-t) & 2\,exp(-t) \\
          -exp(-t)+exp(-2t) & exp(-2t) \\
        \end{array}
      \right).
\end{equation}

In this case, for an arbitrary instant $t_k$ we know that \cite{Troch} $t_k \in (t_{k-1}, \infty)$
and so on.

Thus, choosing $T_{k-l} = 0.8, \;T_k = 1.1$, the expression (25) can be
computed as follows.

According to (31) $\Delta = 0.24743, \; \Delta_1=0.12719, \; \Delta_2 = -0.02014$.

Consequently,
$f_1 =0.51407$ and $f_2= -0.08142$.
According to (10), the functions $g_j^{rq}$ will be
\begin{equation}\label{eq:36}
\left(
  \begin{array}{cccc}
    g_0^{11} & g_1^{11} & g_0^{12} & g_1^{12} \\
    g_0^{21} & g_1^{21} & g_0^{22} & g_1^{22} \\
  \end{array}
\right)
= \left(
    \begin{array}{cccc}
      1. & -0.18121 & 2. & -0.36241 \\
      0. & -0.22207 & 1. & -0.40327 \\
    \end{array}
  \right)
\end{equation}

Substituting the previous expressions, we write the input/output
relations.
\begin{equation*}\label{eq:37}
\left(
\begin{array}{c}
  y_k^1 \\
  y_k^2 \\
\end{array}
\right)
=0.51407\left(
   \begin{array}{c}
     y_{k-1}^1 \\
     y_{k-1}^2 \\
   \end{array}
 \right)
 - 0.08142 \left(
             \begin{array}{c}
               y_{k-2}^1 \\
               y_{k-2}^2 \\
             \end{array}
           \right) +
\end{equation*}

\begin{equation}\label{eq:38}
\left(
    \begin{array}{cccc}
      1. & -0.18121 & 2. & -0.36241 \\
      0. & -0.22207 & 1. & -0.40327 \\
    \end{array}
  \right)
  \left(
    \begin{array}{c}
      u_k^1 \\
      u_{k-1}^1 \\
      u_k^2 \\
      u_{k-1}^2 \\
    \end{array}
  \right).
\end{equation}

For each new sampling instant, the functions $f_i$, $g_j^{rq}$ must be computed
again. The lengths of the sampling intervals can be chosen in order to
optimize a particular performance criterion. Further, difficulties may
arise in the practical implementation of equidistant sampling as, e.g., the
equidistance might be disturbed. The formulation developed above can be
used to pursue the propagation and consequences of this inexactitude.

Finally, it should be mentioned that the use of well-known numerical
methods for the problem of the optimization of aperiodic sampling
instants leads to good results in concrete cases as it may be seen in \cite{Nicoletti}-
\cite{Troch}. The formulation proposed allows us to use these methods for an I/O
modeling technique.

\section{Conclusions}
The external description developed provides a system model well
adapted to the real experimentation conditions although presents the
limitations inherent to the use of the transfer function.

The formulation considered emphasizes the importance of the sampling
sequence against other system parameters.

The particularization to the periodic case is immediate and represents
an alternative to the classic discretization methods without using the \textit{Z}-transform.

The procedure is simple and no use of polynomial matrix theory is
required.

This I/O modeling technique allows us to choose the sampling instants
in order to improve the numerical aspects in problems such as
identification, control, propagation of measuring errors, $\ldots$, etc. Its use
is merely a question of an appropriate choice of the performance criterion.

\end{document}